\documentclass[letterpaper,journal,twocolumn]{IEEEtran}
\usepackage{cite}
\usepackage{amsmath,amssymb,amsthm,amsfonts}
\usepackage{algorithm,algpseudocode}
\usepackage{graphicx}
\usepackage{booktabs,threeparttable}
\usepackage{multirow}
\usepackage{xcolor}
\usepackage{hyperref}
\usepackage{stfloats}
\usepackage[nolist]{acronym}
\usepackage{bm}
\usepackage{bbm}
\usepackage{enumitem}
\hypersetup{hidelinks,
	colorlinks=true,
	allcolors=black,
	pdfstartview=Fit,
	breaklinks=true}

\DeclareMathOperator*{\argmax}{arg\,max}
\newcommand{\PM}{{\rm PM}}
\newcommand{\Idx}[1]{[\![#1]\!]}
\newcommand{\s}[1]{\!#1\!}
\newcommand{\UpRoman}[1]{\MakeUppercase{\romannumeral #1}}
\newcommand{\Rmnum}[1]{\uppercase\expandafter{\romannumeral #1}}

\newtheorem{remark}{Remark}

\begin{document}

\title{Polar Decoding Tree Pruning Based on \\ Soft Output Extraction}

\author{
	\mbox{Li Shen,~\IEEEmembership{Graduate Student Member, IEEE},}
	\mbox{Yongpeng Wu,~\IEEEmembership{Senior Member, IEEE},}
	and \mbox{Wenjun Zhang,~\IEEEmembership{Fellow, IEEE}}
\thanks{The work of Yongpeng Wu was supported in part by the Fundamental Research Funds for the Central Universities, the Yangtze River Delta Science and Technology Innovation Community Joint Research (Basic Research) Project under Grant BK20244006, 111 project BP0719010, and STCSM 22DZ2229005. \emph{(Corresponding author: Yongpeng Wu.)}}
\thanks{Li Shen, Yongpeng Wu, and Wenjun Zhang are with the School of Integrated Circuits (School of Information Science and Electronic Engineering), Shanghai Jiao Tong University, Shanghai 200240, China (e-mail: \href{mailto:shen-l@sjtu.edu.cn}{\mbox{shen-l@sjtu.edu.cn}}; \href{mailto:yongpeng.wu@sjtu.edu.cn}{\mbox{yongpeng.wu@sjtu.edu.cn}}; \href{mailto:zhangwenjun@sjtu.edu.cn}{\mbox{zhangwenjun@sjtu.edu.cn}}).}
}

\begin{acronym}
	\acro{AWGN}{additive white Gaussian noise}
	\acro{B-DMC}{binary-input discrete memoryless channel}
	\acro{BLER}{block error rate}
	\acro{CRC}{cyclic redundancy check}
	\acro{FIC}{frozen-information cascaded}
	\acro{FSCL}{fast SCL}
	\acro{LC-PSCL}{low-complexity partitioned SCL}
	\acro{LDPC}{low-density parity-check}
	\acro{LER}{list error rate}
	\acro{LLR}{log-likelihood ratio}
	\acro{ML}{maximum likelihood}
	\acro{NR}{new radio}
	\acro{PER}{path error rate}
	\acro{PM}{path metric}
	\acro{Rate0}{rate-zero}
	\acro{Rate1}{rate-one}
	\acro{REP}{repetition}
	\acro{SC}{successive cancellation}
	\acro{SCL}{successive cancellation list}
	\acro{SNR}{signal-to-noise ratio}
	\acro{SO-SCL}{soft-output SCL}
	\acro{SO-FSCL}{soft-output FSCL}
	\acro{SOP-SCL}{soft-output-based pruned SCL}
	\acro{SPC}{single-parity-check}
\end{acronym}

\maketitle

% !TEX root = ../main.tex
\begin{abstract}
Although the successive cancellation list (SCL) decoding of polar codes exhibits excellent performance, it retains many decoding paths in the list with negligible contribution to the final output, resulting in high sorting and computational complexity. In this letter, we propose a novel pruning strategy to mitigate the decoding complexity. By leveraging the blockwise soft output extraction process of soft-output SCL and soft-output fast SCL decoding, we provide an accurate approximation of the probability that a decoding path is correct, and thus accordingly prune the paths failing to meet a pre-defined reliability threshold. The complexity reduction achieved by the proposed soft-output-based pruned SCL (SOP-SCL) decoder and its fast version, SOP-FSCL decoder, is significant, without any compromise in error-correction performance. Meanwhile, they also prove to be more efficient than state-of-the-art pruned polar decoders.
\end{abstract}


\section{Introduction}
Among the numerous channel coding schemes, polar codes represent a landmark theoretical breakthrough \cite{Arikan2009Channel}. By performing a breadth-first search within the decoding tree and maintaining the $L$ most probable paths, the \ac{SCL} decoder \cite{Tal2015List} enables polar codes, particularly when combined with \ac{CRC}, to exhibit excellent performance that surpasses Turbo and \ac{LDPC} codes at short-to-medium block lengths \cite{Tal2015List, Niu2012CRC, Balats2015LLR}. However, the superiority of \ac{SCL} is accompanied by a high implementation cost. Firstly, the decoders based on \ac{SC} require bit-by-bit processing, which results in a low degree of parallelism and, consequently, high decoding latency. Moreover, the computational complexity and storage requirements of \ac{SCL} heavily dependent on the list size $L$, such as $L$ times the internal bit and \ac{LLR} updates, $2L$ \ac{PM} calculations, and length-$2L$ \ac{PM} sorting operations. These factors pose significant challenges to the widespread deployment of \ac{SCL} in future latency-sensitive and power-constrained scenarios \cite{Rowshan2024Channel, wu2024physical}.

To mitigate the high latency and complexity of \ac{SCL} decoding, extensive efforts have been devoted. Observe that there exist some special nodes (sub-polar codes) within the recursive structure of polar codes, such as \ac{Rate0}, \ac{Rate1}, and more generalized types. Some studies have proposed \ac{FSCL} decoders by performing path splitting directly on the codeword side of these nodes, instead of traversing each underlying bit \cite{Hashemi2017Fast, Ardakani2019Fast, Ren2022Sequence, Lu2025Fast, yao2023balanced}. Furthermore, during the \ac{SCL} decoding process, many paths have extremely low reliability and contribute negligibly to the final correct decoding. Maintaining these paths in the list causes a significant waste of computational and storage resources. Thus, some researchers have considered identifying and discarding ineffective paths while ensuring that the correct path remains within the list \cite{chen2013reduced, chen2016reduce, zhang2016split, gao2019path, wang2021improved, moradi2023tree, yao2025low}. For example, the erroneous paths in \cite{moradi2023tree} are discarded based on the deviation between the \ac{PM} and the bit-channel mutual information. 
A two-stage pruning strategy is developed in \cite{yao2025low} to eliminate highly unreliable paths and maintain a minimum reliable list of paths by utilizing the evaluated path and list reliability. 

Recently, a \ac{SO-SCL} decoder was proposed in \cite{Yuan2025SoSCL}, which introduced a method to estimate the so-called \textit{codebook probability}. This probability enables \ac{SO-SCL} to generate blockwise soft outputs \cite{Yuan2025SoSCL}, i.e., the probability that a single codeword decision is correct or a candidate list contains the correct codeword. Subsequently, the node-based \ac{FSCL} decoding was introduced into \ac{SO-SCL}. The resulting \ac{SO-FSCL} decoding \cite{Shen2025Soft} significantly reduces decoding latency while ensuring soft-output accuracy.

In this letter, inspired by \cite{Yuan2025SoSCL}, we propose a \ac{SOP-SCL} decoding algorithm and its fast variant, SOP-FSCL decoding. We demonstrate that the reliability of an ongoing decoding path can be evaluated by just employing the intermediate variable used to calculate codebook probability at the current bit level during \ac{SO-SCL}/\ac{SO-FSCL} decoding. Leveraging the evaluated reliabilities, we introduce a threshold-based pruning strategy, which can significantly reduce the average number of decoding paths in the list with almost no error-correction performance loss.

\section{Preliminaries}
\subsection{Notations}
Uppercase and lowercase letters, e.g. $X$ and $x$, represent random variables and their specific realizations, respectively. Bold symbols denote vectors/matrices, and calligraphic fonts denote sets. A length-$N$ vector is written as $\bm{x}^N = (x[1], \dots, x[N])$, where the $i$-th element is given by $x[i]$. The integer set $\{i, i\s{+}1, \dots, j\}$ is abbreviated as $\Idx{i\s{:}j}$, and $\Idx{1\s{:}j}$ is further abbreviated as $\Idx{j}$. For any index set $\mathcal{A} \subseteq \Idx{N}$, the notation $\bm{x}[\mathcal{A}]$ refers to the subvector containing elements $x[i]$ for all $i \in \mathcal{A}$. Specifically, $\bm{x}[\Idx{i\s{:}j}]$ is simply written as $\bm{x}[i\s{:}j]$. Moreover, $\mathcal{X} \setminus \mathcal{Y}$ denotes the set difference between $\mathcal{X}$ and $\mathcal{Y}$, whereas $|\mathcal{X}|$ represents the cardinality of the set $\mathcal{X}$. Given a probability $P$, its approximation is denoted by $P^*$.

\subsection{Polar Codes and Their Decoding Trees}
An $(N, K)$ polar code with block length $N=2^n$ and dimension $K$ is determined by the information index set $\mathcal{I}$ (where $|\mathcal{I}| = K$) and frozen index set $\mathcal{F} = \Idx{N} \setminus \mathcal{I}$. Given the input vector $\bm{u}^N$, $\bm{u}[\mathcal{I}]$ is placed with $K$ information bits, while $\bm{u}[\mathcal{F}]$ is filled with pre-defined frozen bits. Then, the codeword $\bm{c}^N$ of such $(N, K)$ polar code is generated by $\bm{c}^N = \bm{u}^N \bm{G}_N$, where $\bm{G}_N$ is the $n$-th Kronecker power of $\left[\begin{smallmatrix}1&0\\1&1\end{smallmatrix}\right]$ \cite{Arikan2009Channel}.

\begin{figure}[!t]
    \centering
    \includegraphics[width=0.44\textwidth]{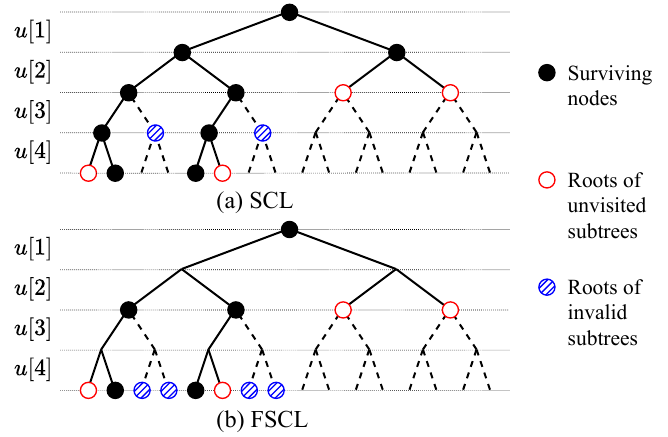}
    \caption{Examples of (a) SCL and (b) FSCL decoding trees of a (4, 3) polar code with frozen bit $u[3] = 0$ and list size $L = 2$. The FSCL decoder only identifies length-2 special subcodes. Thus, $\bm{u}[1\s{:}2]$ is a \ac{Rate1} subcode and $\bm{u}[3\s{:}4]$ is a repetition subcode. Dashed lines indicate pruned branches. This illustration is inspired by \cite[Fig. 1]{Yuan2025SoSCL}.}
    \label{fig:DecTrees}
    \vspace{-10pt}
\end{figure}

At the receiver, the \ac{SC} decoder sequentially estimates each bit $u[i]$ based on the channel observation $\bm{y}^N$ and previously determinations $\hat{\bm{u}}[1\s{:}i\s{-}1]$ as \cite{Arikan2009Channel}
\begin{equation}
    \hat{u}[i] \s{=} \left\{\begin{aligned}
    &\hspace{-0.05cm} \,\text{frozen value}, && \hspace{-0.2cm} i\s{\in}\mathcal{F}; \\
    &\hspace{-0.05cm} \argmax_{b\in\{0,1\}} P_{\bm{Y}^N, \bm{U}^{i\s{-}1}|U[i]}\left(\bm{y}^N, \hat{\bm{u}}[1\s{:}i\s{-}1]|b\right), && \hspace{-0.2cm} i\s{\in}\mathcal{I}.
    \end{aligned}\right. \hspace{-0.2cm}
\end{equation}
Furthermore, the \ac{SCL} decoder evaluates both hypotheses of $u[i]$, $i\in\mathcal{I}$, to double the decoding paths, and only retains the $L$ paths with the minimum \acp{PM}. For a decoding path $\bm{a}_l^{i}$ with path index $l$, its \ac{PM} $\PM^{(l)}_{i}$is associated with the path probability $P_{\bm{U}^i|\bm{Y}^N}$ by \cite{Balats2015LLR}
\begin{equation}
    \ln P_{\bm{U}^i|\bm{Y}^N} (\bm{a}_l^{i}|\bm{y}^N) = -\PM^{(l)}_{i}.
\end{equation}

The \ac{SCL} decoding process can be represented by a binary tree shown in Fig. \ref{fig:DecTrees}(a), where each node at the $i$-th level (the tree roots at the 0-th level) corresponds to a potential decoding path $\bm{a}^i \in \{0,1\}^i$. Assuming that the surviving nodes at the $(i\s{-}1)$-th level are contained in the set $\mathcal{V}_{i\s{-}1}$, descending to the next level will categorizes their children into three sets \cite{Yuan2025SoSCL}: 1) $\mathcal{V}_i$, containing the still-surviving nodes; 2) $\mathcal{W}_i$, containing the valid nodes discarded due to the list size limit, whose descendant subtrees are unvisited; and 3) the set of invalid nodes that violate the frozen constraints. 

If $\bm{u}[i_s\s{:}j_s]$ form a special sub-polar code in the recursive manner of polar codes, the \ac{FSCL} decoder can obtain the estimates of $\bm{u}[i_s\s{:}j_s]$ simultaneously rather than bit-by-bit processing \cite{Hashemi2017Fast, Ardakani2019Fast, Ren2022Sequence, Lu2025Fast, yao2023balanced}. That is, \ac{FSCL} will bypass the intermediate nodes and directly perform path splitting and selection (i.e., the survival and  discarding of nodes) at the $j_s$-th level in the decoding tree, as depicted in Fig. \ref{fig:DecTrees}(b). We denote such a subcode as $\mathbb{N}_{i_s}^{j_s}$, and the set of $\mathbb{N}_{i_s}^{j_s}$ constituting an $(N, K)$ polar code as $\mathcal{N}$ hereafter.

\subsection{SO-SCL Decoding}
The \ac{SO-SCL} decoder aims to leverage the \ac{SCL} decoding tree to estimate the codebook probability $P_\mathcal{U}(\bm{y}^N)$ as
\begin{equation}
    P_\mathcal{U}(\bm{y}^N) \triangleq \sum_{\bm{u}^N \in \mathcal{U}} P_{\bm{U}^N|\bm{Y}^N} \left(\bm{u}^N|\bm{y}^N\right),
\end{equation}
where $\mathcal{U}$ contains all valid input vectors $\bm{u}^N$ that satisfy the frozen constraints \cite{Yuan2025SoSCL}. However, the paths in $\mathcal{T} = \mathcal{U}\setminus\mathcal{V}_N$, inherited from the nodes in $\cup_{i\in\Idx{N}} \mathcal{W}_i$, are discarded by \ac{SCL}, and thus their path probabilities remain unknown.

To address this, \ac{SO-SCL} assumes that the frozen bits are uniformly distributed, and approximates the term $P_\mathcal{T}(\bm{y}^N) = \sum_{\bm{u}^N \in \mathcal{T}} P_{\bm{U}^N|\bm{Y}^N} \left(\bm{u}^N|\bm{y}^N\right)$ by\footnote{Indeed, \cite{Yuan2025SoSCL} shows that this approximation remains sufficiently accurate for static frozen bits.}
\begin{equation}
    P^*_{\mathcal{T}}(\bm{y}^N) = \sum_{i \in \Idx{N}} \sum_{\bm{a}^i \in \mathcal{W}_i} 2^{-\left|\mathcal{F}^{(i:N)}\right|} P_{\bm{U}^i|\bm{Y}^N}\left(\bm{a}^i|\bm{y}^N\right),
    \label{eq:Pt_SCL}
\end{equation}
where $\mathcal{F}^{(i:j)} = \{k \mid k\in\mathcal{F}, i<k\leq j\}$ \cite{Yuan2025SoSCL}. In this way, the codebook probability $P_\mathcal{U}(\bm{y}^N)$ can be approximated as
\begin{equation}
	P^*_{\mathcal{U}}(\bm{y}^N) = \sum_{\bm{u}^N\in \mathcal{V}_N} P_{\bm{U}^N|\bm{Y}^N}\left(\bm{u}^N|\bm{y}^N\right) + P^*_{\mathcal{T}}(\bm{y}^N).
\end{equation}

Relying on $P^*_\mathcal{U}(\bm{y}^N)$, \ac{SO-SCL} can extract blockwise soft outputs \cite{Yuan2025SoSCL}. Specifically, it approximates the probability that a single decision $\hat{\bm{u}}^N$ is correct by
\begin{equation}
    \Gamma^*\left(\hat{\bm{u}}^N, \bm{y}^N\right) = \frac{P_{\bm{U}^N|\bm{Y}^N}\left(\hat{\bm{u}}^N|\bm{y}^N\right)}{P^*_{\mathcal{U}}(\bm{y}^N)},
    \label{eq:PathSO}
\end{equation}
and further approximates the probability that a candidate list $\mathcal{L}$ contains the correct codeword by
\begin{equation}
    \Gamma^*\left(\mathcal{L}, \bm{y}^N\right) = \sum_{\bm{u}^N \in \mathcal{L}} \Gamma^*\left(\bm{u}^N, \bm{y}^N\right).
    \label{eq:ListSO}
\end{equation}

\subsection{SO-FSCL Decoding}
By applying \ac{FSCL} decoding, the calculation of $P^*_{\mathcal{T}}(\bm{y}^N)$ relates to the types of $\mathbb{N}_{i_s}^{j_s}$ and (\ref{eq:Pt_SCL}) is then rewritten as
\begin{equation}
    P^*_{\mathcal{T}}(\bm{y}^N) = \sum_{\mathbb{N}_{i_s}^{j_s} \in \mathcal{N}} 2^{-\left|\mathcal{F}^{(j_s:N)}\right|} P_{\mathcal{W}_{j_s}}(\mathbb{N}_{i_s}^{j_s}),
    \vspace{-0.1cm}
\end{equation}
where $P_{\mathcal{W}_{j_s}}(\mathbb{N}_{i_s}^{j_s})$ is defined according to (\ref{eq:Pt_SCL}) as
\begin{equation}
    P_{\mathcal{W}_{j_s}}(\mathbb{N}_{i_s}^{j_s}) = \sum_{\bm{a}^{j_s} \in \mathcal{W}_{j_s}} P_{\bm{U}^{j_s}|\bm{Y}^N}\left(\bm{a}^{j_s}|\bm{y}^N\right).
    \label{eq:Pwj}
\end{equation}

The \ac{SO-FSCL} decoder in \cite{Shen2025Soft} identifies \ac{FIC} subcodes\footnote{As for why the \ac{FIC} constraint is adopted and the impact on the calculation of $P_{\mathcal{W}_{j_s}}(\mathbb{N}_{i_s}^{j_s})$ when it is not satisfied, please referred to \cite[Remark 2]{Shen2025Soft}.}, i.e., $\Idx{i_s\s{:}f_s} \subseteq \mathcal{F}$, where $f_s = i_s\s{+}F_s\s{-}1$ and $F_s$ is the number of frozen bits in $\mathbb{N}_{i_s}^{j_s}$. If $N_s-F_s \leq \min\{3, N_s/2\}$, where $N_s = j_s-i_s+1$, $P_{\mathcal{W}_{j_s}}(\mathbb{N}_{i_s}^{j_s})$ could be directly calculated by (\ref{eq:Pwj}). Otherwise, since \ac{FSCL} may not access the probabilities of all nodes in $\mathcal{W}_{j_s}$, $P_{\mathcal{W}_{j_s}}(\mathbb{N}_{i_s}^{j_s})$ should be calculated in an alternative way by (\ref{eq:Pwj_H}), as shown at the bottom of this page. Specifically, if $F_s > 0$, the probability $P_{\bm{U}^{f_s}|\bm{Y}^N}$ is obtained via an auxiliary \ac{SCL} decoder that only decodes frozen bits. Then, the soft output extraction is consistent with that of \ac{SO-SCL}.
\stepcounter{equation}

\section{Soft-Output-Based Pruning for SCL and FSCL Decoding}
\subsection{Truncated Codebook Probability}
\begin{figure}[!t]
    \centering
    \includegraphics[width=0.47\textwidth]{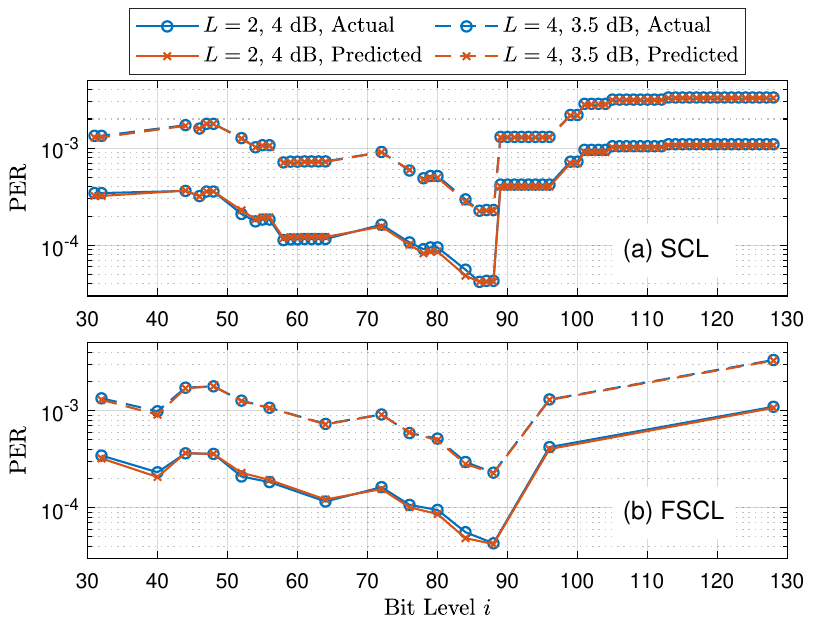}
    \caption{Comparison of actual and predicted PERs at different bit levels for the $(128, 64)$ 5G polar code with (a) SCL and (b) FSCL decoding.}
    \label{fig:PER}
    \vspace{-10pt}
\end{figure}

The codebook probability estimates the sum of probabilities for all valid codewords given channel observation $\bm{y}^N$. Through (\ref{eq:PathSO}) and (\ref{eq:ListSO}), it enables the reliability evaluation of a single or a list of candidate codewords. Along this line, to evaluate the reliability of an ongoing decoding path, we should estimate the probability sum of all valid paths at the current decoding level. This sum, which we refer to as the truncated codebook probability, is defined as
\begin{equation}
    P_{\mathcal{U}_i}(\bm{y}^N) \triangleq \sum_{\bm{u}^i \in \mathcal{U}_i} P_{\bm{U}^i|\bm{Y}^N} \left(\bm{u}^i|\bm{y}^N\right),
\end{equation}
where $\mathcal{U}_i = \{\bm{u}[1\s{:}i] \mid \bm{u}^N \in \mathcal{U}\}$ and $i \in \Idx{N}$. Similarly, since the paths in $\mathcal{T}_i = \mathcal{U}_i\setminus\mathcal{V}_i$ are discarded by \ac{SCL}/\ac{FSCL} decoding, we need to approximate the term $P_{\mathcal{T}_i}(\bm{y}^N) =$ $\sum_{\bm{u}^i \in \mathcal{T}_i} P_{\bm{U}^i|\bm{Y}^N} \left(\bm{u}^i|\bm{y}^N\right)$. In the following, we will elaborate on how to leverage the soft output extraction processes to approximate $P_{\mathcal{T}_i}(\bm{y}^N)$ and subsequently approximate $P_{\mathcal{U}_i}(\bm{y}^N)$ for \ac{SCL} and \ac{FSCL} decoding\footnote{To facilitate soft output extraction, the \ac{FSCL} decoding throughout this letter identifies the same special subcodes as the \ac{SO-FSCL} decoding.}, respectively.

\begin{figure*}[b]
\vspace{-10pt} \hrulefill
\begin{equation}
    P_{\mathcal{W}_{j_s}}(\mathbb{N}_{i_s}^{j_s}) = \sum_{\bm{a}^{i_s\s{-}1} \in \mathcal{V}_{i_s\s{-}1}} P_{\bm{U}^{f_s}|\bm{Y}^N} \left((\bm{a}^{i_s\s{-}1}, \bm{0}^{F_s})|\bm{y}^N\right) - \sum_{\bm{a}^{j_s} \in \mathcal{V}_{j_s}} P_{\bm{U}^{j_s}|\bm{Y}^N} \left(\bm{a}^{j_s}|\bm{y}^N\right),
    \tag{10}
    \label{eq:Pwj_H}
\end{equation}
\begin{equation}
    P^*_{\mathcal{U}_{j_s}}(\bm{y}^N) = \begin{cases}
    2^{-F_s} P^*_{\mathcal{T}_{i_s-1}}(\bm{y}^N) + \sum\limits_{\bm{a}^{j_s} \in \mathcal{V}_{j_s}\cup\mathcal{W}_{j_s}} P_{\bm{U}^{j_s}|\bm{Y}^N}\left(\bm{a}^{j_s}|\bm{y}^N\right), &N_s-F_s \leq \min\{3, N_s/2\}; \\
    2^{-F_s} P^*_{\mathcal{T}_{i_s-1}}(\bm{y}^N) + \sum\limits_{\bm{a}^{i_s\s{-}1} \in \mathcal{V}_{i_s\s{-}1}} P_{\bm{U}^{f_s}|\bm{Y}^N} \left((\bm{a}^{i_s\s{-}1}, \bm{0}^{F_s})|\bm{y}^N\right), &N_s-F_s > \min\{3, N_s/2\}.
    \end{cases}
    \tag{16}
    \label{eq:Pui_FSCL}
\end{equation}
\vspace{-10pt}
\end{figure*}

\subsubsection{SCL Decoding}
Based on the assumption of uniformly distributed frozen bits in \ac{SO-SCL} \cite{Yuan2025SoSCL}, $P_{\mathcal{T}_i}(\bm{y}^N)$ can be approximated as in (\ref{eq:Pt_SCL}) by
\begin{equation}
    P^*_{\mathcal{T}_i}(\bm{y}^N) = \sum_{k \in \Idx{i}} \sum_{\bm{a}^k \in \mathcal{W}_k} 2^{-\left|\mathcal{F}^{(k:i)}\right|} P_{\bm{U}^k|\bm{Y}^N}\left(\bm{a}^k|\bm{y}^N\right).
    \label{eq:Pti_SCL}
\end{equation}
Furthermore, we express (\ref{eq:Pti_SCL}) into a recursive update form as 
\begin{equation}
\begin{aligned}
    P^*_{\mathcal{T}_i}(\bm{y}^N) &= 2^{-\mathbbm{1}_\mathcal{F}(i)} P^*_{\mathcal{T}_{i-1}}(\bm{y}^N) + \sum_{\bm{a}^i \in \mathcal{W}_i} P_{\bm{U}^i|\bm{Y}^N}\left(\bm{a}^i|\bm{y}^N\right) \\
    &= 
    \begin{cases}
        P^*_{\mathcal{T}_{i-1}}(\bm{y}^N)/2, & i\in\mathcal{F}; \\
        P^*_{\mathcal{T}_{i-1}}(\bm{y}^N) + \hspace{-0.1cm} \sum\limits_{\bm{a}^i \in \mathcal{W}_i} \hspace{-0.2cm} P_{\bm{U}^i|\bm{Y}^N}\left(\bm{a}^i|\bm{y}^N\right), & i\in\mathcal{I},\\ 
    \end{cases}
\end{aligned}
\label{eq:Pti_SCL_r}
\end{equation}
where $\mathbbm{1}_\mathcal{F}(i)$ is the indicator function which equals 1 if $i \in \mathcal{F}$ and 0 otherwise, and the second equality is based on the observation $\mathcal{W}_i = \emptyset$ for $i \in \mathcal{F}$. Thus, the truncated codebook probability $P_{\mathcal{U}_i}(\bm{y}^N)$ is approximated as
\begin{equation}
\begin{aligned}
    P^*_{\mathcal{U}_i}(\bm{y}^N) &= \sum_{\bm{a}^i\in \mathcal{V}_i} P_{\bm{a}^i|\bm{Y}^N}\left(\bm{a}^i|\bm{y}^N\right) + P^*_{\mathcal{T}_i}(\bm{y}^N) \\
    &= 
    \begin{cases}
        \sum\limits_{\bm{a}^i \in \mathcal{V}_i} \hspace{-0.1cm} P_{\bm{a}^i|\bm{Y}^N}\left(\bm{a}^i|\bm{y}^N\right) + P^*_{\mathcal{T}_{i-1}}(\bm{y}^N)/2, \; i\in\mathcal{F}; \\
        \hspace{-0.3cm} \sum\limits_{\hspace{0.3cm} \bm{a}^{i\s{-}1} \in \mathcal{V}_{i\s{-}1}} \hspace{-0.5cm} P_{\bm{a}^{i\s{-}1}|\bm{Y}^N}\left(\bm{a}^{i\s{-}1}|\bm{y}^N\right) + P^*_{\mathcal{T}_{i-1}}(\bm{y}^N), i\in\mathcal{I}, \\
    \end{cases} \hspace{-0.3cm}
\end{aligned}
\end{equation}
where the case of $i \in \mathcal{I}$ in the second equality is derived from the fact that the path splitting leads to $\mathcal{V}_i \cup \mathcal{W}_i = \{(\bm{a}^{i\s{-}1}, b) \mid \bm{a}^{i\s{-}1} \in \mathcal{V}_{i\s{-}1}, b \in \{0,1\}\}$.

\subsubsection{FSCL Decoding}
For \ac{FSCL} decoding, the approximation of $P_{\mathcal{T}_i}(\bm{y}^N)$ shares a similar approach with that of \ac{SCL} decoding in (\ref{eq:Pti_SCL}). Following the recursive update of $P^*_{\mathcal{T}_i}(\bm{y}^N)$ in (\ref{eq:Pti_SCL_r}), after decoding each special subcode $\mathbb{N}_{i_s}^{j_s}$, $P^*_{\mathcal{T}_{j_s}}(\bm{y}^N)$ at the \mbox{$j_s$-th} level is updated as
\begin{equation}
    P^*_{\mathcal{T}_{j_s}}(\bm{y}^N) = 2^{-F_s} P^*_{\mathcal{T}_{i_s-1}}(\bm{y}^N) + P_{\mathcal{W}_{j_s}}(\mathbb{N}_{i_s}^{j_s}).
    \label{eq:Pti_FSCL}
\end{equation}
Then, by substituting (\ref{eq:Pwj}) or (\ref{eq:Pwj_H}) into (\ref{eq:Pti_FSCL}), we can use $P^*_{\mathcal{T}_{j_s}}(\bm{y}^N)$ to approximate the corresponding truncated codebook probability $P_{\mathcal{U}_{j_s}}(\bm{y}^N)$ by (\ref{eq:Pui_FSCL}) at the bottom of this page.
\stepcounter{equation}

It is worth noting that $P^*_{\mathcal{T}_N}(\bm{y}^N)$ is equivalent to $P^*_{\mathcal{T}}(\bm{y}^N)$ at the $N$-th level. This implies that updating $P^*_{\mathcal{T}_i}(\bm{y}^N)$ not only enables the calculation of $P_{\mathcal{U}_i}(\bm{y}^N)$ to support the path reliability evaluation and pruning in subsequent subsections, but also yields the approximated codebook probability $P^*_{\mathcal{U}}(\bm{y}^N)$ required by \ac{SO-SCL}/\ac{SO-FSCL}.

\subsection{Path Reliability}
Given an ongoing decoding path $\bm{a}^i$ of depth $i$, we can approximate the probability of this path being correct, analogously to (\ref{eq:PathSO}), by
\begin{equation}
    \Gamma^*_i\left(\bm{a}^i, \bm{y}^N\right) = \frac{P_{\bm{U}^i|\bm{Y}^N}\left(\bm{a}^i|\bm{y}^N\right)}{P^*_{\mathcal{U}_i}(\bm{y}^N)}.
    \label{eq:PathSOi}
\end{equation}
To assess this approximation, we compare the \ac{PER} of the path with the minimum \ac{PM} in the list at $i$-th level against its \ac{PER} predicted by $\mathbb{E}[1-\Gamma^*_i\left(\bm{a}^i, \bm{y}^N\right)]$. \mbox{Fig. \ref{fig:PER}} presents such a comparison for decoding the $(128, 64)$ 5G polar code under different simulation settings. As observed, the predicted \ac{PER} tightly matches the actual \ac{PER}, which indicates that the approximation $\Gamma^*_i\left(\bm{a}^i, \bm{y}^N\right)$ can serve as a reasonable metric for path reliability.

\subsection{Pruning Strategy}
Based on the evaluation of path reliability, we introduce the following strategy to prune highly unreliable paths that are unlikely to survive in the subsequent list. 

A decoding path $\bm{a}^i$ is considered highly unreliable if it satisfies
\begin{equation}
    \Gamma^*_i\left(\bm{a}^i, \bm{y}^N\right) \leq \eta_P,
    \label{eq:PathP}
\end{equation}
where $\eta_P$ denotes the threshold for unreliability tolerance. Given that at a certain \ac{SNR}, the error probability of final decision after \ac{SCL}/\ac{FSCL} decoding is $\varepsilon$, i.e., the \ac{BLER} is $\varepsilon$, we propose relating the threshold $\eta_P$ to $\varepsilon$ by
\begin{equation}
    \eta_P = \delta \times \varepsilon.
\end{equation}
Here, $\delta$ is a scaling factor for tightening the threshold to avoid degradation in error-correction performance, since a path with $\Gamma^*_i\left(\bm{a}^i, \bm{y}^N\right)$ slightly below $\varepsilon$ might still be correct with a non-negligible probability.

Then, after path splitting, we prune all paths in the list that satisfy (\ref{eq:PathP}). Note that this pruning operation is independent of the \ac{PM} ordering. Hence, it can be executed prior to \ac{PM} sorting to alleviate the sorting overhead.

In summary, during each path splitting and selection phase at $i$-th level, we perform the following steps:
\begin{enumerate}[label=\arabic*.]
    \item Split the decoding paths to obtain $\mathcal{V}^{\prime\prime}_i$ and corresponding \acp{PM}, followed by the calculation of $P^*_{\mathcal{U}_i}(\bm{y}^N)$\footnote{\label{fn:algo} The \ac{FSCL} decoder may perform multiple path splitting and selection operations at the same level, while $P^*_{\mathcal{U}_i}(\bm{y}^N)$ only needs to be calculated at the first time and $P^*_{\mathcal{T}_{i}}(\bm{y}^N)$ only needs to be updated at the last time.};
    \item Prune the paths in $\mathcal{V}^{\prime\prime}_i$ that satisfy (\ref{eq:PathP}) to obtain $\mathcal{V}^{\prime}_i$;
    \item Sort the paths in $\mathcal{V}^{\prime}_i$ descendingly by their \acp{PM} to obtain the ordered list $\bar{\mathcal{V}}^{\prime}_i$;
    \item Retain the first $\min\{|\bar{\mathcal{V}}^{\prime}_i|, L\}$ paths in $\bar{\mathcal{V}}^{\prime}_i$ to obtain the final surviving list $\mathcal{V}_i$ and update $P^*_{\mathcal{T}_{i}}(\bm{y}^N)$\textsuperscript{\ref{fn:algo}}.
\end{enumerate}

The standard \ac{SCL} and \ac{FSCL} decoding equipped with the proposed pruning strategy are termed the \ac{SOP-SCL} and SOP-FSCL decoding algorithms, respectively. Compared with \ac{SCL}/\ac{FSCL}, the additional operations in \ac{SOP-SCL}/SOP-FSCL lie in updating $P^*_{\mathcal{T}_{i}}(\bm{y}^N)$ and $P^*_{\mathcal{U}_i}(\bm{y}^N)$, and pruning using $P^*_{\mathcal{U}_i}(\bm{y}^N)$. To facilitate practical implementation, we may follow \cite[Sec. \UpRoman{3}-E]{Shen2025Soft} to calculate $P^*_{\mathcal{T}_{i}}(\bm{y}^N)$ and $P^*_{\mathcal{U}_i}(\bm{y}^N)$ in the log domain and adopt a hardware-friendly version. Thus, both the path reliability evaluation in (\ref{eq:PathSOi}) and pruning decision in (\ref{eq:PathP}) could be easily implemented in the log domain.

\begin{remark}
    It is also possible to evaluate the reliability for an ongoing path list $\mathcal{L}_i$ by $\Gamma^*_i\left(\mathcal{L}_i, \bm{y}^N\right) = \sum_{\bm{a}^i \in \mathcal{L}_i} \Gamma^*_i\left(\bm{a}^i, \bm{y}^N\right)$ as in (\ref{eq:ListSO}), and subsequently keep only the minimum list that meets the reliability threshold post-sorting (akin to the selection step in \cite{yao2025low}). However, we found through simulations that this strategy has a restricted effect on reducing the average number of paths. Moreover, if our pruning strategy is executed first, it will yield negligible further reduction. The underlying reason may be that our strategy already preserves a near-minimal list containing the correct path by filtering out unreliable paths precisely. Hence, this approach is not included in our strategy.
\end{remark}

\section{Simulation Results}
In this section, we evaluate the \ac{BLER} performance and the corresponding decoding complexity of the proposed SOP-SCL and SOP-FSCL decoders over \ac{AWGN} channels. The decoding complexity primarily stems from sorting and computation. Specifically, the sorting complexity is measured by the average number of paths sent for sorting per codeword, while the computational complexity is evaluated by the average unit calculations, counted as the number of executing the $f_-$ and $f_+$ functions \cite[Eq. (8)]{Balats2015LLR}. 

\begin{figure}[!t]
    \centering
    \includegraphics[width=0.47\textwidth]{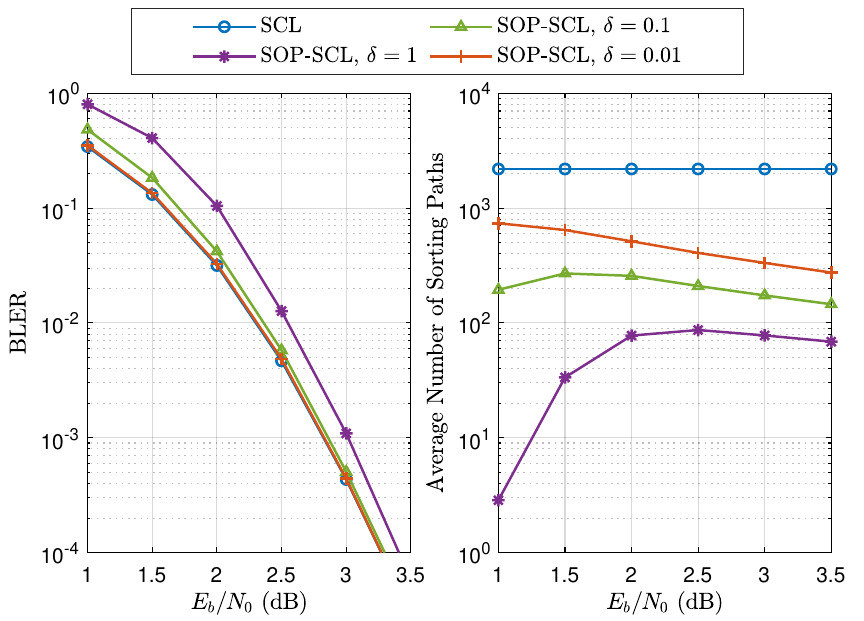}
    \caption{BLER performance and sorting complexity of SOP-SCL decoding for (256, 128) 5G polar code with $L=8$ and CRC-11 under different factors $\delta$.}
    \label{fig:Delta}
    \vspace{-10pt}
\end{figure}

We first discuss the determination of the scaling factor $\delta$. Fig. \ref{fig:Delta} illustrates the impact of different $\delta$ on the \ac{BLER} performance and sorting complexity. It can be observed that a trade-off exists between performance and complexity. Although a tighter threshold ensures no degradation in error-correction performance, it also retains more ineffective paths, resulting in a increase in complexity. Meanwhile, at higher \acp{SNR}, since the probability difference between incorrect and correct paths becomes more pronounced, the performance degradation under the same $\delta$ is also smaller. In this work, we simply choose the parameter $\delta$ as $\delta = 10^{-k}$ with the largest possible integer $k$ that does not cause any performance loss at a target \ac{BLER} of $10^{-2}$. The resulting $\delta$ is then used over the entire \ac{SNR} range.

\begin{figure}[!t]
    \centering
    \includegraphics[width=0.47\textwidth]{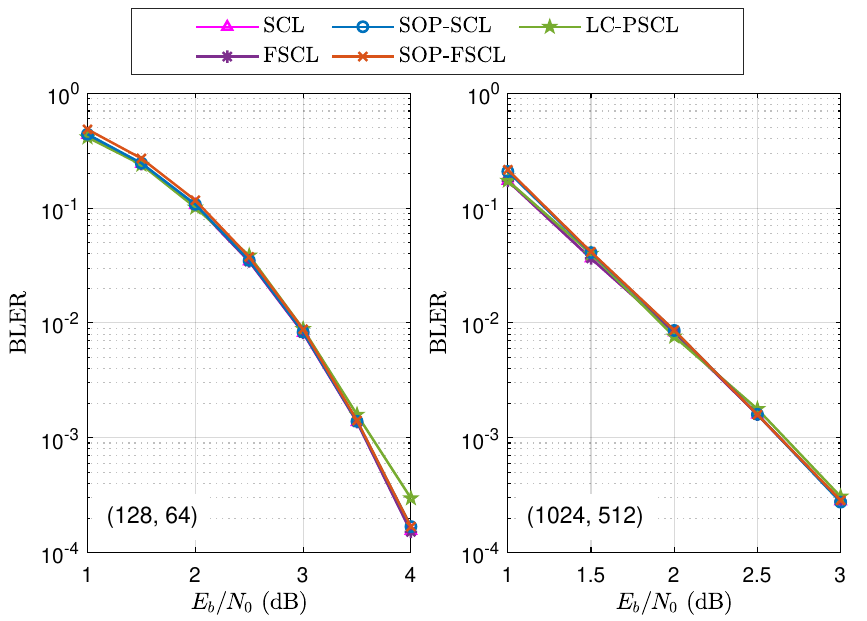}
    \includegraphics[width=0.47\textwidth]{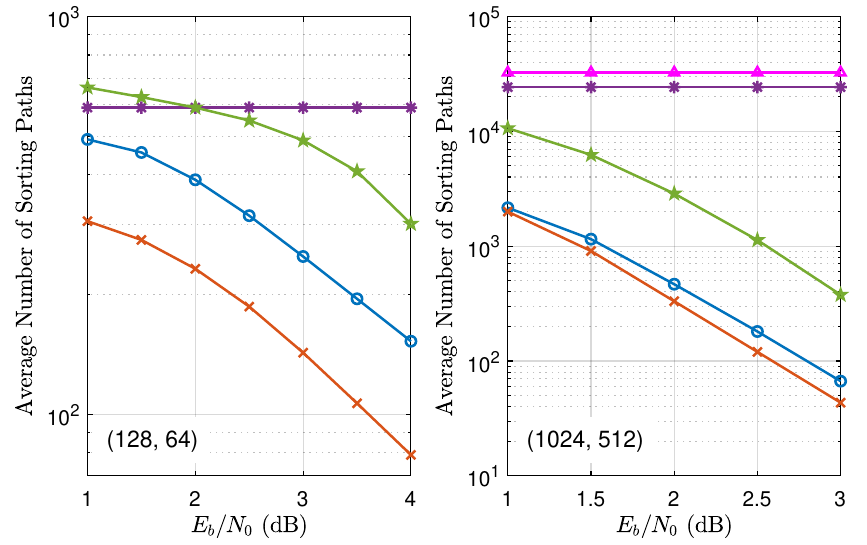}
    \includegraphics[width=0.47\textwidth]{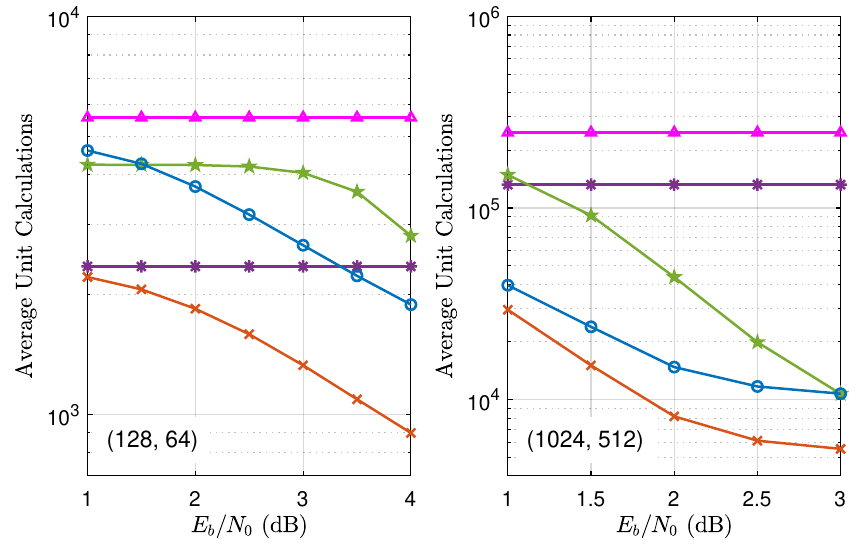}
    \caption{BLER performance, sorting complexity, and computational complexity of different decoders for (128, 64) 5G polar code with CRC-11 and $L=8$, and (1024, 512) 5G polar code with CRC-0 and $L=32$.}
    \label{fig:Code}
    \vspace{-10pt}
\end{figure}

Furthermore, we evaluate the \ac{BLER} performance, sorting complexity, and computational complexity of (128, 64) and (1024, 512) 5G polar codes under different decoders in \mbox{Fig. \ref{fig:Code}}. The results of the \ac{LC-PSCL} proposed in \cite{yao2025low} are also presented. For both \ac{SOP-SCL} and SOP-FSCL, $\delta$ is set to 0.01. We can observe that compared with conventional \ac{SCL} decoding, the proposed \ac{SOP-SCL} can significantly reduce decoding complexity without \ac{BLER} performance loss, which can be further reduced by adopting its fast version. Particularly for the (1024, 512) code, the complexity saving ratio of SOP-FSCL even exceeds 97\%. Meanwhile, by leveraging a more accurate evaluation of path reliability, the proposed SOP-SCL/SOP-FSCL is also more efficient than LC-PSCL.

\section{Conclusion}
This letter proposes a polar decoding tree pruning strategy based on soft output extraction. Leveraging the SO-SCL/SO-FSCL decoding process, the truncated codebook probability is approximated, allowing for an accurate evaluation of the decoding path reliability. Subsequently, by pruning those unreliable paths falling below the threshold, the proposed SOP-SCL and SOP-FSCL decoders substantially decrease the decoding complexity while not compromising BLER performance.

Although SOP-SCL/SOP-FSCL are still capable of providing soft outputs like SO-SCL/SO-FSCL, this letter focuses more on their advantage as hard-output decoders in reducing the complexity of SCL/FSCL. The application of SOP-SCL/SOP-FSCL in iterative detection and iterative decoding scenarios, as in \cite{Yuan2025SoSCL,Shen2025Soft}, requires further investigation. In addition, exploring more fine-grained settings for the pruning thresholds (e.g., at the bit level) and providing corresponding theoretical guidance is another direction.

\bibliographystyle{IEEEtran}
\bibliography{refer}

@article{Shen2025Soft,
  title = {Node-based soft-output fast successive cancellation list decoding of polar codes},
  author = {Shen, Li and Wu, Yongpeng and Gao, Zhen and Xu, Yin and You, Xiaohu and Gao, Xiqi and Zhang, Wenjun},
  journal={{IEEE} Trans. Commun.},
  volume={74},
  pages={8500--8516},
  month={May},
  year={2026},
}

@article{Arikan2009Channel,
  title={Channel polarization: A method for constructing capacity-achieving codes for symmetric binary-input memoryless channels},
  author={Ar{\i}kan, Erdal},
  journal={{IEEE} Trans. Inf. Theory},
  volume={55},
  number={7},
  pages={3051--3073},
  month={Jul.},
  year={2009},
}

@article{Tal2015List,
  title={List decoding of polar codes},
  author={Tal, Ido and Vardy, Alexander},
  journal={{IEEE} Trans. Inf. Theory},
  volume={61},
  number={5},
  pages={2213--2226},
  month={May},
  year={2015},
}

@article{Niu2012CRC,
  title={{CRC}-aided decoding of polar codes},
  author={Niu, Kai and Chen, Kai},
  journal={{IEEE} Commun. Lett.},
  volume={16},
  number={10},
  pages={1668--1671},
  month={Sep.},
  year={2012},
}

@article{Balats2015LLR,
  title={{LLR}-based successive cancellation list decoding of polar codes},
  author={Balatsoukas-Stimming, Alexios and Parizi, Mani Bastani and Burg, Andreas},
  journal={{IEEE} Trans. Signal Process.},
  volume={63},
  number={19},
  pages={5165--5179},
  month={Oct.},
  year={2015},
}

@article{Rowshan2024Channel,
  title={Channel coding toward {6G}: Technical overview and outlook}, 
  author={Rowshan, Mohammad and Qiu, Min and Xie, Yixuan and Gu, Xinyi and Yuan, Jinhong},
  journal={{IEEE} Open J. Commun. Soc.}, 
  volume={5},
  pages={2585--2685},
  month={Apr.},
  year={2024},
}

@article{wu2024physical,
  title={Physical layer signal processing for {XR} communications and systems},
  author={Wu, Yongpeng and Xu, Mai and Zhai, Guangtao and Zhang, Wenjun},
  journal={Sci. China Inf. Sci.},
  volume={67},
  number={12},
  month={Nov.},
  year={2024, Art. no. 221301},
}

@article{Hashemi2017Fast,
  title={Fast and flexible successive-cancellation list decoders for polar codes},
  author={Hashemi, Seyyed Ali and Condo, Carlo and Gross, Warren J},
  journal={{IEEE} Trans. Signal Process.},
  volume={65},
  number={21},
  pages={5756--5769},
  month={Nov.},
  year={2017},
}

@article{Ardakani2019Fast,
  title={Fast successive-cancellation-based decoders of polar codes},
  author={Ardakani, Maryam Haghighi and Hanif, Muhammad and Ardakani, Masoud and Tellambura, Chintha},
  journal={{IEEE} Trans. Commun.},
  volume={67},
  number={7},
  pages={4562--4574},
  month={Jul.},
  year={2019},
}

@article{Ren2022Sequence,
  title={A sequence repetition node-based successive cancellation list decoder for {5G} polar codes: Algorithm and implementation},
  author={Ren, Yuqing and Kristensen, Andreas Toftegaard and Shen, Yifei and Balatsoukas-Stimming, Alexios and Zhang, Chuan and Burg, Andreas},
  journal={{IEEE} Trans. Signal Process.},
  volume={70},
  pages={5592--5607},
  year={2022},
}

@article{Lu2025Fast,
  title={Fast list decoding of high-rate polar codes},
  author={Lu, Yang and Zhao, Ming-Min and Lei, Ming and Zhao, Min-Jian},
  journal={{IEEE} Trans. Commun.}, 
  volume={73},
  number={1},
  pages={22--38},
  month={Jan.},
  year={2025},
}

@article{yao2023balanced,
  title={A balanced tree approach to construction of length-flexible polar codes},
  author={Yao, Xinyuanmeng and Ma, Xiao},
  journal={{IEEE} Trans. Commun.},
  volume={72},
  number={2},
  pages={665--674},
  month={Feb.},
  year={2024},
}

@inproceedings{chen2013reduced,
  title={A reduced-complexity successive cancellation list decoding of polar codes},
  author={Chen, Kai and Niu, Kai and Lin, Jiaru},
  booktitle={{IEEE} Veh. Technol. Conf. (VTC Spring)},
  pages={1--5},
  month={Jun.},
  year={2013},
  address={Dresden, Germany},
}

@article{chen2016reduce,
  title={Reduce the complexity of list decoding of polar codes by tree-pruning},
  author={Chen, Kai and Li, Bin and Shen, Hui and Jin, Jie and Tse, David},
  journal={{IEEE} Commun. Lett.},
  volume={20},
  number={2},
  pages={204--207},
  month={Feb.},
  year={2016},
}

@article{zhang2016split,
  title={A split-reduced successive cancellation list decoder for polar codes},
  author={Zhang, Zhaoyang and Zhang, Liang and Wang, Xianbin and Zhong, Caijun and Poor, H Vincent},
  journal={{IEEE} J. Sel. Areas Commun.},
  volume={34},
  number={2},
  pages={292--302},
  month={Feb.},
  year={2016},
}

@article{gao2019path,
  title={Path splitting selecting strategy-aided successive cancellation list algorithm for polar codes},
  author={Gao, Chenyu and Liu, Rongke and Dai, Bin and Han, Xu},
  journal={{IEEE} Commun. Lett.},
  volume={23},
  number={3},
  pages={422--425},
  month={Mar},
  year={2019},
}

@article{wang2021improved,
  title={An improved path splitting decision-aided {SCL} decoding algorithm for polar codes},
  author={Wang, Xiumin and Zhang, Hongchao and Li, Jun and Bao, Xiupin and Xie, Kunyu},
  journal={{IEEE} Commun. Lett.},
  volume={25},
  number={11},
  pages={3463--3467},
  month={Nov.},
  year={2021},
}

@article{moradi2023tree,
  title={A tree pruning technique for decoding complexity reduction of polar codes and {PAC} codes},
  author={Moradi, Mohsen and Mozammel, Amir},
  journal={{IEEE} Trans. Commun.},
  volume={71},
  number={5},
  pages={2576--2586},
  month={May},
  year={2023},
}

@article{yao2025low,
  title={Low-complexity {PSCL} decoding of polar codes},
  author={Yao, Xinyuanmeng and Ma, Xiao},
  journal={{IEEE} Trans. Commun.},
  volume={73},
  number={9},
  pages={7021--7031},
  month={Sep.},
  year={2025},
}

@ARTICLE{Yuan2025SoSCL,
  title={Soft-output successive cancellation list decoding}, 
  author={Yuan, Peihong and Duffy, Ken R. and Médard, Muriel},
  journal={{IEEE} Trans. Inf. Theory}, 
  volume={71},
  number={2},
  pages={1007--1017},
  month = {Feb.},
  year={2025},
}

\end{document}